# Coherence of Spin Qubits in Silicon

A. M. Tyryshkin[1], J. J. L. Morton[2,3], S. C. Benjamin[2],
A. Ardavan[3], G. A. D. Briggs[2], J. W. Ager[4], and S. A. Lyon[1]

[1]Department of Electrical Engineering, Princeton, NJ 08544
[2]Department of Materials, Oxford University, Parks Road, Oxford OX1 3PH, UK
[3]The Clarendon Laboratory, Oxford University, Parks Road, Oxford OX1 3PU, UK
[4]Materials Sciences Division, Lawrence Berkley National Laboratory, Berkeley, CA 94720

Email: lyon@princeton.edu

**Abstract.** Given the effectiveness of semiconductor devices for classical computation one is naturally led to consider semiconductor systems for solid state quantum information processing. Semiconductors are particularly suitable where local control of electric fields and charge transport are required. Conventional semiconductor electronics is built upon these capabilities and has demonstrated scaling to large complicated arrays of interconnected devices. However, the requirements for a quantum computer are very different from those for classical computation, and it is not immediately obvious how best to build one in a semiconductor. One possible approach is to use spins as qubits: of nuclei, of electrons, or both in combination. Long qubit coherence times are a prerequisite for quantum computing, and in this paper we will discuss measurements of spin coherence in silicon. The results are encouraging – both electrons bound to donors and the donor nuclei exhibit low decoherence under the right circumstances. Doped silicon thus appears to pass the first test on the road to a quantum computer.

## 1. Introduction

Semiconductor based qubits (quantum bits) and gates were among the earliest suggestions for physical realizations of quantum information processors [1-4]. Since those early proposals, numerous groups have tackled various aspects of the problem of defining and constructing quantum logic in a semiconductor. The popularity of semiconducting systems for quantum computers can be directly traced to their popularity for classical electronics. A huge base of knowledge and experience has been built up over the last half-century about all aspects of semiconductors – their chemical purification, crystal growth, defect control, nanostructure fabrication, and so forth. Given their versatility, many different states inside a semiconductor have been proposed as qubits. Some of these include excitons bound in quantum dots [1,2,5,6], the spin of electrons trapped at donors or in quantum dots [4,7-10], other low-lying states of impurities [11], and nuclear spins of an either an impurity [3,12] or the host semiconductor [13,14], as well as combinations of one or more of these states.

In this paper we will limit ourselves to spin qubits, and consider both electrons and nuclei. Even with these limitations there are still numerous approaches to constructing the various quantum gates which are required for information processing. We will also limit ourselves to a single host semiconductor, silicon, because it is particularly well-suited to obtaining long spin coherence. Among the common semiconductors, only the elemental ones (C, Si, and Ge) have stable isotopes without nuclear magnetic moments. The presence of unwanted magnetic moments can lead to decoherence of nuclear spins through spin diffusion. Such processes will also decohere the electron spins, as will motion of an electron through the random magnetic landscape of the nuclear moments. Recent work has demonstrated ways of mitigating some of the ill effects of the nuclear moments in other semiconductors [15,16], but spin coherence times are not yet comparable to those in Si. Among the elemental semiconductors, germanium has a relatively strong spin-orbit interaction, while the spin-orbit effect for X-point electrons in silicon and diamond is unexpectedly weak [17]. However, silicon device technology is much more advanced than that of diamond (though without some of the unique possibilities, such as fullerenes and nanotubes), making it reasonably straightforward to envision quantum devices based upon well-known classical structures.

Shallow donors in silicon have many features which make them attractive candidates for spin qubits in a quantum computer [3,7,18-20]. The ground electronic state of the donors (except Li) is symmetrical and spin-degenerate only, with a large gap (> 10 meV) to the excited states, leading to exceptionally long electron spin relaxation times at liquid helium temperatures. The first measurements of electron spin relaxation for donors in silicon were done in late 1950's by Honig [21] and by Feher and Gere [22]. They found unusually long relaxation times ($T_{1e}$) which at first were assigned to nuclear relaxation processes [23] but later were confirmed to describe electron spin relaxation. Relaxation times $T_{1e}$ ~ 1 hour were measured at 1.25 K, limited by the direct one-phonon mechanism [24] when great care was taken to prevent exposure of the sample to light or room temperature radiation [22]. At higher temperatures (2-20 K) a two-phonon Orbach mechanism [25] begins to be important and the relaxation times drop sharply by many orders of magnitude to ~$10^{-6}$ s at 20 K [22,26]. The transverse relaxation time was also found to be long, though considerably shorter than the longitudinal times; $T_{2e}$ = 240 μs in natural silicon and $T_{2e}$ = 520 μs in isotopically-purified $^{28}$Si [27,28]. Recently it has been found that longer transverse times are possible: $T_{2e}$ = 60ms was determined at 7K in isotopically-purified $^{28}$Si as will be discussed further below [29]. It is expected that $T_{2e}$ will be longer at lower temperatures and with further isotopic purification of the $^{28}$Si. Spin relaxation of the donor nuclei have been measured at high doping densities by nuclear magnetic resonance (NMR), but there is insufficient sensitivity to extend those measurements to much lower densities [30]. We will discuss recent measurements of the coupled system of electron and nuclear spins of a neutral $^{31}$P donor in $^{28}$Si using electron nuclear double resonance (ENDOR) [31].

## 2. Experimental Details

The silicon crystals used in these experiments were doped with phosphorus in the range $10^{15}$ - $10^{16}$ P/cm$^3$. All other electrically active impurities had concentrations < mid-$10^{14}$ cm$^{-3}$. Both natural silicon crystals (4.7% of $^{29}$Si) and isotopically-purified $^{28}$Si, either bulk grown or as epitaxial layers 10-25 μm thick on p-type natural Si were used. The $^{28}$Si-enriched material contained a residual $^{29}$Si concentration of ~800 ppm as determined by secondary ion mass spectrometry [32,33]. Pulsed EPR experiments were done using an X-band Bruker EPR spectrometer (Elexsys 580) equipped with a low temperature helium-flow cryostat (Oxford CF935). The temperature was controlled with a precision of better than 0.05 K using calibrated temperature sensors (Lakeshore Cernox CX-1050-SD) and an Oxford ITC503 temperature controller. This precision was needed because of the strong temperature dependence of the electron spin relaxation times (e.g., $T_{1e}$ varies by 5 orders of magnitude between 7 K and 20 K). The electron spin $T_{2e}$ and $T_{1e}$ were measured using 2-pulse electron spin echo (ESE) and inversion recovery experiments, respectively [34]. In the *bang-bang* decoupling experiment [35], a modified Davies ENDOR sequence was used with a second refocusing RF pulse added at the end of the pulse sequence [36]. Microwave pulses of duration 16 and 32 ns were used for π/2 and π rotations of the electron spin and RF pulses of 15-100 μs were used for π rotation of the $^{31}$P nuclear spins.

## 3. Electron Spin Relaxation for Shallow Donors in Silicon

*3.1 Spin relaxation due to $^{29}$Si nuclei in natural Si:P*

Natural silicon contains 4.7% of $^{29}$Si with q nuclear spin I = 1/2. Spins of two neighboring $^{29}$Si nuclei can flip-flop to exchange their polarization and by this means the spin polarization can travel from one $^{29}$Si site to another through the lattice, termed nuclear spin diffusion. An electron spin residing on a donor interacts with the surrounding $^{29}$Si nuclei through the hyperfine interactions (contact and dipole) and therefore feels the $^{29}$Si nuclear spin diffusion as fluctuations in the local magnetic field. This fluctuating nuclear field results in additional dephasing of the electron spin; a mechanism known as spectral diffusion.

The theory of nuclear-induced spectral diffusion was developed in 1960-70's [37-39] and more recently adapted to the exact wave-functions and the lattice structure of P donors in silicon [40]. It results in non-exponential spin relaxation decays which can be described by $V(\tau) = V_0 \exp\left[-\left(2\tau/T_{SD}\right)^n\right]$, where τ is the time between the two pulses of a Hahn echo experiment, $T_{SD}$ is the characteristic time of spectral diffusion and n is an exponent stretching factor which may vary between 2 and 3 for different regimes of spectral diffusion [39]. Both $T_{SD}$ and n are complicated functions of the nuclear spin concentration and the relative position of the flip-flopping nuclei with respect to each other and also of the flip-flopping pair with respect to the electron spin in the host lattice [40]. The breadth of the electron spin wavefunction and its magnitude on the surrounding nuclei is an important factor in determining the transition between different spectral diffusion regimes, e.g. from n ~ 3 to n ~ 2 [39].

To experimentally characterize the role of $^{29}$Si-induced spectral diffusion in donor spin relaxation, we performed Hahn echo experiments using natural silicon with low P doping (8·10$^{14}$ P/cm$^3$). At this low doping, the instantaneous diffusion effects (see next section) are not excessive and therefore the effect of nuclear spectral diffusion can be observed most cleanly. The measured echo decays (Figure 1A) are non-exponential and also strongly dependent on the orientation of the applied magnetic field with respect to the crystal axes. The longest decay is found for the magnetic field oriented along [100] and the shortest decay for the field oriented along [111]. We fit these non-exponential decays using a function of the form:

$$V(\tau) = V_0 \exp\left[-2\tau/T_{2e} - \left(2\tau/T_{SD}\right)^n\right], \quad (1)$$

where, in addition to the spectral diffusion term we include a second exponential time constant $T_{2e}$ to account for other relaxation processes, including the $T_{1e}$ related processes and instantaneous diffusion, which can be described by a simple exponential decay. In our fit we assumed $T_{2e}$ to be orientation independent and the fit resulted in $T_{2e} = 1.1$ ms, mostly limited by instantaneous diffusion at this doping density. Both $T_{SD}$ and n obtained from the fit show a strong orientation dependence (Figure 1B), with $T_{SD}$ changing from 0.62 ms for the field oriented along the [100]-direction to 0.27 ms along [111], and with n changing from 3 along [100] to 2.4 for the field tilted by $\theta \geq 20^\circ$ from [100].

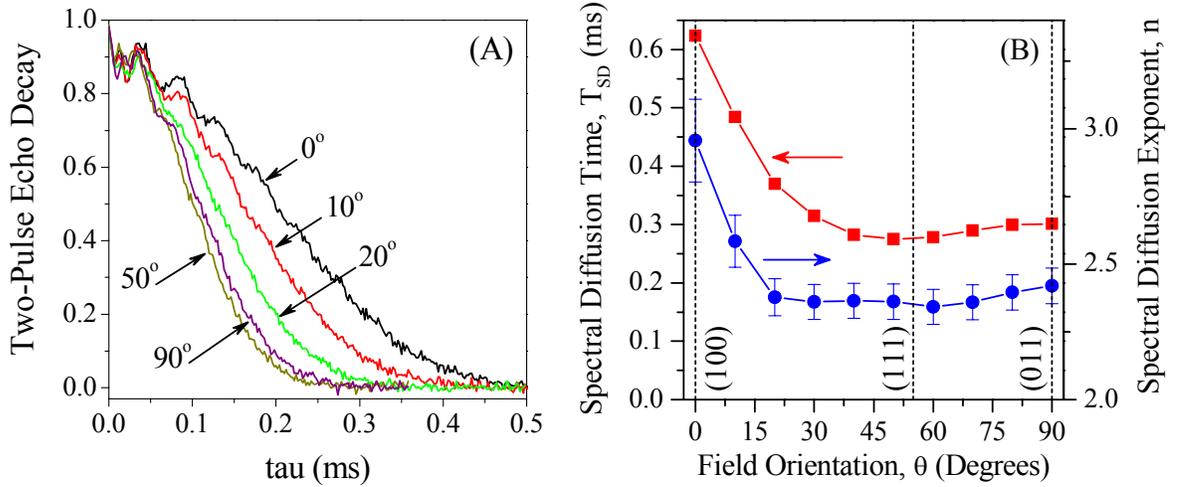

Figure 1. (A) The two-pulse echo decays in natural Si:P ($8 \cdot 10^{14}$ P/cm$^3$) at 8 K for selected orientations of the applied magnetic field $B_0$ with respect to the [100] crystal axis (field orientation is indicated with angle in degrees for each curve). The crystal rotation is done in the [100]-[011] plane. Weak oscillations superimposed on the decays at small $\tau$ are the electron spin echo envelope modulation caused by $^{29}$Si. (B) The orientation dependence of the spectral diffusion time, $T_{SD}$ (■, left axis) and the exponent, n (●, right axis) extracted from fitting the decay using Eqn. (1). The vertical dashed lines indicate field orientations along specific axes of the silicon crystal.

Gordon and Bowers noted that the echo decays were non-exponential [27] and Chiba and Hirai [28] have quantified these non-exponential decays for Si:P in echo experiments at a single field orientation along the [111]-direction. Their $T_{SD} = 0.36$ ms is in reasonable agreement with our data, however their n = 3 is noticeably different from our n = 2.4 along [111]. The non-exponential decays and similar orientation dependence for $T_{SD}$ has also been reported by Abe et al. [41] in natural silicon as well as in $^{29}$Si-enriched silicon. However, they found an orientation independent n, and assumed n ~ 2 at all orientations in their analysis. For $^{29}$Si-enriched silicon (~99% $^{29}$Si), they found $T_{SD}$ to be about an order of magnitude shorter than in natural silicon, as would be expected because of faster spin diffusion in the $^{29}$Si-rich media. In the next section we demonstrate that in $^{28}$Si-purified silicon the non-exponential part in the echo decay is strongly suppressed and thus long purely-exponential decays are observed (we estimate $T_{SD} > 8$ ms for 800 ppm of residual $^{29}$Si in our $^{28}$Si-purified silicon).

The orientation dependence for $T_{SD}$ in silicon has recently been predicted theoretically by de Souza et al. [40]. Their simulated decays were analyzed in terms of a phase memory time, $T_M$, defined as the time for the echo signal decay to 1/e times its original magnitude. The stretch factor was calculated for a single orientation [111], where n = 3 was found. The trend in the predicted orientation dependence for $T_M$ matches quite well with our $T_{SD}$ dependence in Figure 1B, except for an overall scaling by a factor of 3.

*3.2 Isotopically-Purified $^{28}$Si:P*

In very pure $^{28}$Si silicon (800ppm of residual $^{29}$Si), the effect of nuclear spectral diffusion is small and therefore very long, exponential echo decays can be observed. While measuring long two-pulse echo decays we faced the problem of phase instability of the echo signal. This is illustrated in Figure 2A where the as measured in-phase and quadrature signals of the microwave detector are shown in a single-shot experiment (eg. without signal averaging). At long interpulse delays τ (> 0.5 ms) strong "noise" starts to develop and dominates the in-phase and quadrature signals. However, this noise largely disappears when the magnitude of the echo signal is calculated (bottom trace in Figure 2A). Apparently the spins remain in phase with one another and form a strong echo signal at long τ, but they go out of phase with the spectrometer microwave source and therefore the echo signal fluctuates between the in-phase and quadrature detection channels. These phase fluctuations are caused by fluctuations in the magnetic field during the two-pulse echo experiment and possibly by fluctuations in the phase of the microwave source. We observed that characteristics of this instrumental noise vary significantly on different days and also depend upon various instrumental settings (orientation of the modulation coils with respect to the main magnetic field, settings in automatic frequency control circuit, etc.).

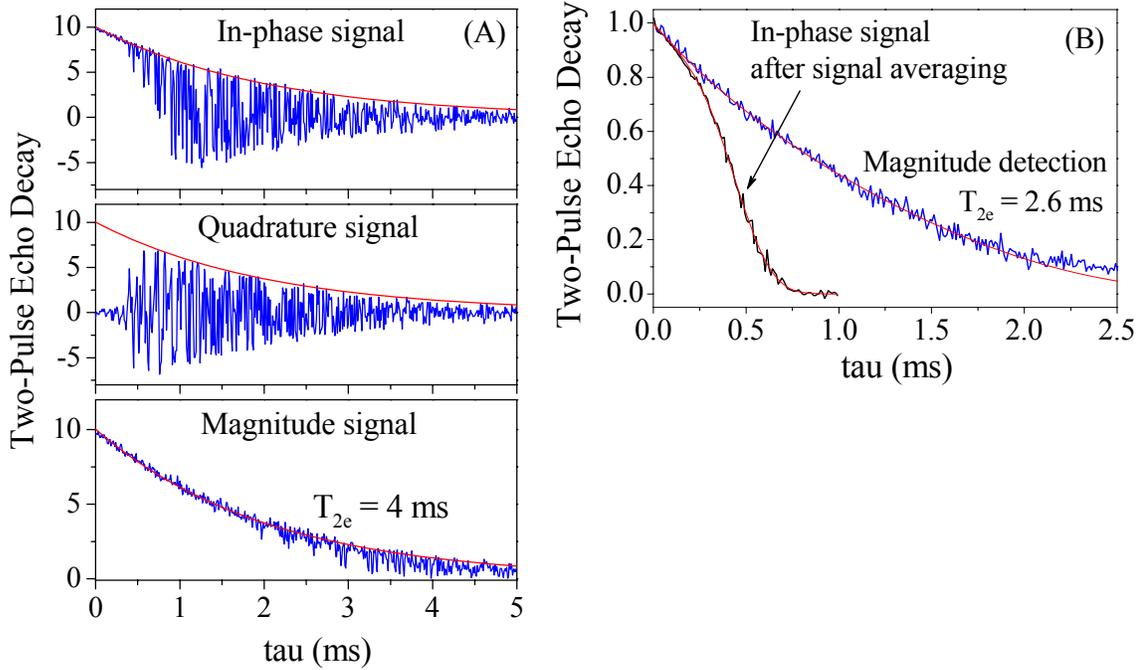

Figure 2. The 2-pulse echo decay for isotopically-purified $^{28}$Si:P crystal (5·10$^{14}$ P/cm$^3$) at 8 K. (A) The in-phase (top) and quadrature (middle) signals measured in a single-shot experiment (eg. no signal averaging). The fluctuations are nearly eliminated when the magnitude of the echo signal (bottom) is calculated as [in-phase$^2$ + quadrature$^2$]$^{1/2}$. The exponential fit (red traces in each plot) corresponds to $T_{2e}$ = 4 ms. (B) Signal averaging of the fluctuating in-phase signal results in a distorted, non-exponential echo decay approximated as $\exp[-2\tau/T_{2e} - (2\tau/T_{inst})^n]$, with $T_{inst}$ =1.05 ms and n = 3.6. Signal averaging of the magnitude signal reveals the much longer, exponential decay with $T_{2e}$ = 2.6 ms for $^{28}$Si:P (9·10$^{14}$ P/cm$^3$) at 7 K.

Because of this instrumental phase noise, repetitive summation of the in-phase and quadrature echo signals (eg. signal averaging to improve signal-to-noise) results in distorted echo decays with strongly non-exponential characteristics (Figure 2B). To avoid these instrumental problems and to detect very long, undistorted echo decays we use single-shot detection. Instead of averaging the two detection channels separately and then obtaining the magnitude, we first calculate the magnitude signal, disregarding the fluctuating phase, and average this signal in repetitive experiments to improve the signal to noise. This use of this magnitude detection approach requires that the signal is strong enough to be detected in a single-shot experiment. This requirement places a severe limit on two-pulse echo experiments, and only those samples with a spin concentration of about 10$^{15}$/cm$^3$ or larger give sufficient signal for this procedure [29].

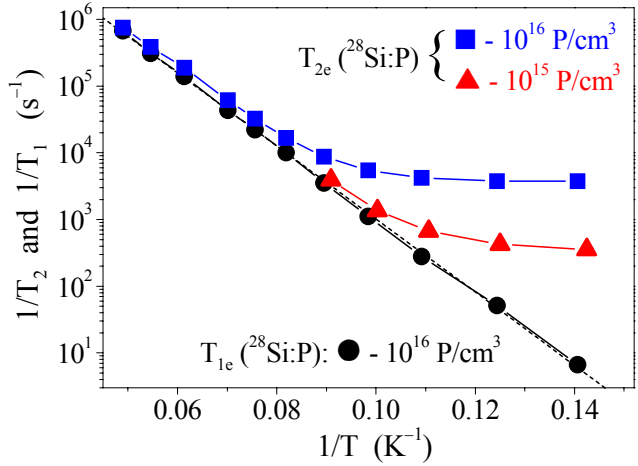

Figure 3. Temperature dependence of electron spin relaxation times, $T_{1e}$ and $T_{2e}$, in isotopically-purified $^{28}$Si:P. The linear dependence of $T_{1e}$ corresponds to the Orbach mechanism controlling the relaxation at this temperature range [26]. $T_{2e}$ is controlled by the $T_{1e}$ processes at high temperatures (T > 10K) and by instantaneous diffusion at low temperatures [29].

The $T_{1e}$ and $T_{2e}$ data and their temperature dependence in the range 7-20 K have been recently measured using pulsed ESR for isotopically-purified $^{28}$Si:P with $10^{15}$ - $10^{16}$ P/cm$^3$ [29]. In this temperature range $T_{1e}$ is found to be independent of both the P concentration and the density of $^{29}$Si. The Arrhenius plot (Figure 3) shows that $T_{1e}$ is controlled by an Orbach relaxation process with an energy gap to the first excited state, $\Delta E = 126$ K. This result is in good agreement with previous conclusions derived from continuous wave ESR measurements [26].

The temperature dependence of $T_{2e}$ is more complex (Figure 3). At high temperatures 12-20 K, $T_{2e}$ closely follows the $T_{1e}$ dependence, and thus $T_{2e}$ is fully controlled by the $T_{1e}$ relaxation processes in this temperature range. However, at lower temperature $T_{2e}$ diverges from $T_{1e}$, and while $T_{1e}$ continues to grow $T_{2e}$ levels off and becomes temperature independent. The fact that the low-temperature $T_{2e}$ is approximately 10 times larger in the sample with a 10 times smaller P concentration suggests that at low temperatures $T_{2e}$ is mostly determined by the dipole-dipole interactions between the donor electron spins.

One aspect of the dipole-dipole interaction between spins and its effect on a two-pulse echo experiment has been termed instantaneous diffusion [42]. The set of two-pulse echo decays shown in Figure 4A was obtained using a variable rotation angle, $\theta_2$, of the second microwave pulse. The relaxation time increases significantly at smaller rotation angles, and a very long $T_{2e} = 14$ ms is found at $\theta_2 = 45°$ as compared to $T_{2e} = 3.1$ ms at $\theta_2 = 170°$. This coherence time of 14 ms is the longest which has been directly measured for P donors in silicon. However, plotting the relaxation rate, $1/T_{2e}$, against $\sin^2(\theta_2/2)$ in Figure 4B reveals a linear dependence with the slope proportional to the P donor concentration. By extrapolating to a very small $\theta_2$ we are able to extract the $T_{2e}$ =

60 (+50/–20) ms which corresponds to the $T_{2e}$ expected for an isolated donor in $^{28}$Si. As also shown in Figure 4B, this extrapolated $T_{2e}$ is similar to but somewhat shorter than $T_{1e}$ measured for P donors at this temperature (6.9K). Thus, $T_{2e}$ of the isolated donor in $^{28}$Si is probably limited by the $T_{1e}$ relaxation processes down to at least 7 K.

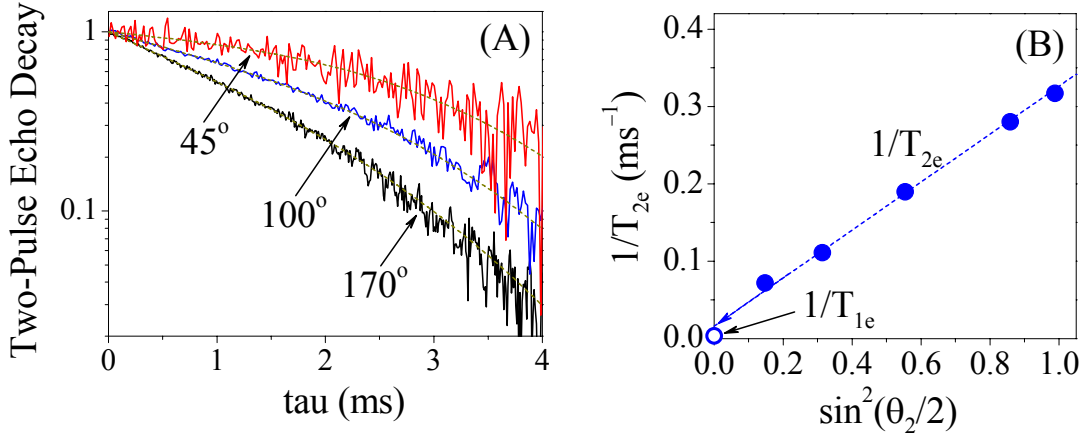

Figure 4. (A) The 2-pulse echo decays for isotopically-purified $^{28}$Si:P (9·10$^{14}$ P/cm$^3$) at 6.9 K and at different rotation angles $\theta_2$ (indicated in degrees for each curve) of the second microwave pulse. The dashed lines are fits using $\exp[-2\tau/T_{2e} - (2\tau/T_{inst})^3]$, where the cubic term $T_{inst} \sim 8$ ms originates from incomplete phase noise cancellation and/or spectral diffusion from residual $^{29}$Si (800 ppm). The fit to the echo decay curve with $\theta_2 = 45°$ gives $T_{2e} = 14$ ms. (B) A plot the relaxation rates $1/T_{2e}$ (●) as a function of the turning angle forms a line which extrapolates to $T_{2e} = 60$ ms at small $\theta_2$. The point (○) corresponds to the measured $T_{1e} = 280$ ms at 6.9 K.

**4. Donor Nuclear Spins in Si:P**

*4.1 Coding Spin Qubits in Silicon Donors*

      All shallow donors in silicon (and their various isotopes) have non-zero nuclear spins and thus quite naturally, both the electron and nuclear spins of neutral donors have been proposed to be used for coding, manipulating and storing quantum information [3,7,18]. The $^{31}$P donors are most popular choice among common donors because both the electron and nucleus have spin 1/2 and thus two qubits can be encoded using electron and nuclear spin states. The energy level diagram and qubit coding scheme are shown in Figure 1A. In the presence of a strong magnetic field the hyperfine coupling between the spins results in a non-uniform spacing of the energy levels and therefore selective excitation (or addressing) of the individual electron and nuclear spin transitions is possible by applying in-resonance microwave and radio frequency (RF) pulses. Single-qubit operations (spin rotations) are implemented by applying two pulses to coherently rotate two resolved electron (or nuclear) spin transitions. The two-qubit CNOT operation is even easier to perform since it requires only one RF pulse. The single-qubit gates and

the two-qubit CNOT gate provide the universal set of gates and thus any other desired gates can be implemented [43].

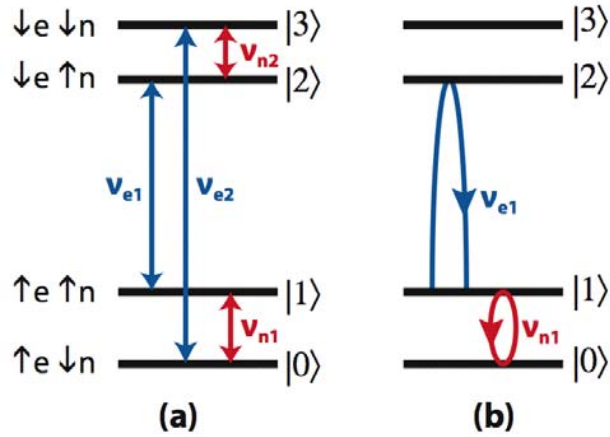

Figure 5. (a) An energy level diagram for the coupled electron-nuclear spin pair of neutral $^{31}$P donor in silicon (electron spin S = 1/2, nuclear spin I = 1/2). In an applied magnetic field ($B_0$ = 0.35 T at X-band ESR), the transitions of the electrons are driven by microwave frequencies ($v_{e1}$, $v_{e2}$) near 10 GHz, while the nuclear spin transitions are driven by radio frequencies ($v_{n1}$, $v_{n2}$) at 50-60 MHz. The hyperfine coupling between the spins forces the spin energy levels to be spaced non-uniformly ($v_{e1} \neq v_{e2}$, and $v_{n1} \neq v_{n2}$) and therefore each of the allowed electron and nuclear transitions can be addressed individually with selective microwave and RF pulses, respectively. (b) In the *bang-bang* decoupling experiment a $2\pi$ rotation ($v_{e1}$) is applied to the electron spin to produce a fast phase shift and thus to refocus the evolution of the nuclear spin driven by field $v_{n1}$ [35].

*4.2 Bang-Bang Decoupling of $^{31}$P Nuclear Spin Using Controlled Flips of Electron Spin*

The ability to preserve quantum information coherently over an extended period of time is a prerequisite for a quantum system to be useful as qubit. As has been discussed above, the spin of electrons bound to donors have coherence times of at least $T_{2e}$ = 60 ms at liquid helium temperatures which permits at least $10^6$ single-qubit operations before the spin decoheres (we assume 60 ns long $2\pi$ pulses available in a standard pulsed ESR spectrometer). Although the relaxation times for donor nuclear spin have not been measured yet, it is anticipated that they are also very long, possibly in excess of the electron spin relaxation times. Longer relaxation times might be expected because of the smaller magnetic moment of the nuclei and thus weaker coupling to the fluctuating environment.

The concept of dynamical decoupling (DD), using a series of fast symmetrizing pulses to reduce (average out) the undesired parts of the system-environment interactions has been developed recently [44,45]. If the DD scheme is introduced on top of the already long relaxation times in Si:P, the coherent evolution period of the system can be extended even further. Here we demonstrate one possible DD implementation for Si:P using the advantage of having two strongly coupled spins, electron and nuclear, in the donor. We implement a *bang-bang* decoupling pulse protocol to manipulate one (electron) spin in the coupled pair to effectively decouple the second (nuclear) spin from the decohering environment. This approach was first demonstrated for a similar system of coupled electron and nuclear spins (S = 3/2 and I = 1) in endohedral fullerenes, N@$C_{60}$ [35].

To demonstrate the full power of the *bang-bang* decoupling pulse scheme, we intentionally introduce a strong "environmental" perturbation to the nuclear $^{31}$P spin of the donor by applying a resonant RF field ($\nu_{n1}$) to drive Rabi nutations between the nuclear spin states |0> and |1> (Figure 5B). This strong RF field is then successfully decoupled by applying fast, selective $2\pi$ pulses ($\nu_{e1}$) to rotate the electron spin around closed cycles between states |1> and |2>. By mean of these selective $\nu_{e1}$ rotations, rapid 180º phase shifts are introduced to the nuclear spin state |1>, while the phase of state |0> remains unchanged:

$$\Psi_i = a|0\rangle + b|1\rangle \xrightarrow{2\pi(\nu_{e1})} \Psi_f = a|0\rangle - b|1\rangle. \qquad (2)$$

This phase shift can refocus the RF-driven evolution of the nuclear spin.

The experimental demonstration of this effect is shown in Figure 6. The unperturbed Rabi nutation of the nuclear spin between the states |0> and |1> is driven by the long RF pulse ($\nu_{n1}$) as shown in Figure 6A. The amplitude of the Rabi oscillations decreases as the RF pulse duration increases because the RF field inhomogeneity; spins in different part of the sample have slightly different nutation frequency and gradually lose coherence at long times. Application of the $2\pi$ pulse to the donor electron spin at time $t_p$ = 92 μs induces a nearly instantaneous (on the time scale of the nuclear nutation) phase shift to the nuclear spin nutation (Figure 6B). The action of the RF field is reversed following this phase shift. After a further evolution period $t_p$ (at the time indicated with ▲ in Figure 6B) the nuclear spin nutation recovers its full amplitude indicating that all nuclear spins are in phase again and thus the decoherence caused by the inhomogeneity of the RF field is fully refocused at this point.

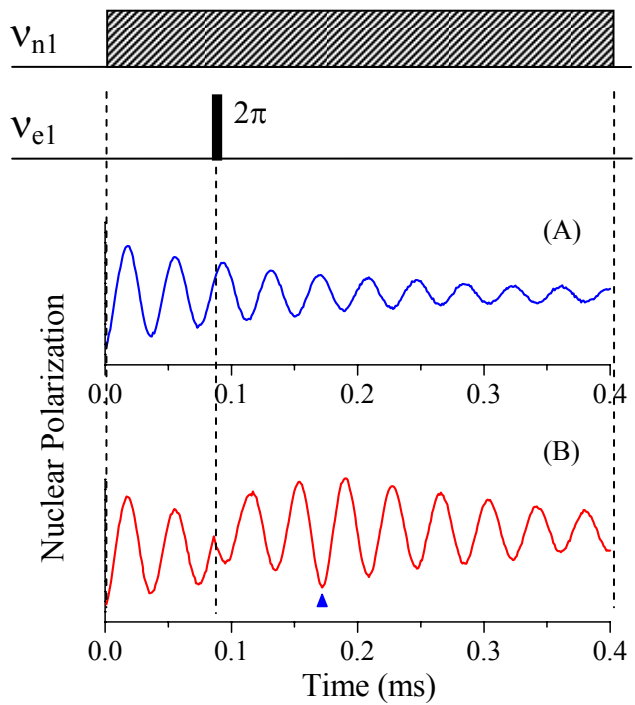

Figure 6. Rabi nutation of the $^{31}$P nuclear spin of phosphorus donor in silicon driven by a long RF pulse at $\nu_{n1}$ = 52.33 MHz. The timing diagram for the RF ($\nu_{n1}$) and microwave ($\nu_{e1}$) pulses used in the experiment is shown at top of the figure (the respective transitions are indicated in Figure 1B). Free Rabi nutation in (A) is interrupted in (B) by applying a fast $2\pi$ rotation ($\nu_{e1}$) to the electron spin at time 92 μs. This $2\pi$ rotation causes a nearly instantaneous phase shift to the nuclear spin and refocuses its evolution back to the initial state (indicated by ▲). The observed baseline shift after application the $2\pi$ pulse is the result of an imperfect $2\pi$ rotation of the electron spin.

Figure 7 shows a logical extension to the above experiment [35]. By applying a series of the $2\pi$ pulses at a higher repetition rate than the nuclear nutation frequency, the nuclear spin evolution can be continuously refocused and thus locked in one particular state (Figure 7B). It can then be released as desired to be locked again later in a different state upon application of a second series of the $2\pi$ pulses (Figure 7C). These experiments demonstrate an unprecedented level of environmental decoupling (even against the strong in-resonance RF field) which can be achieved using a relatively simple *bang-bang* pulse protocol. Our ability to implement this method for the donor in silicon is directly related to having two coupled (electron and nuclear) spins in the donor and the ability to selectively and independently rotate each allowed spin transition in the coupled system. This demonstrates the potential benefits of physical 'qubit' systems beyond the simple 2-level structure.

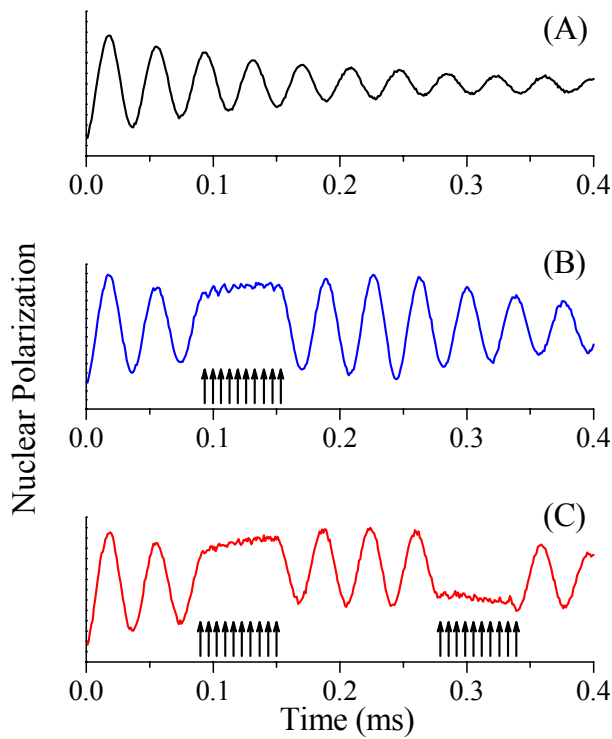

Figure 7. (A) Free Rabi nutation of the $^{31}$P nuclear spin driven by a long RF pulse at $\nu_{n1}$ can be stopped at will and locked into a particular spin state (B) by applying a burst of the closely spaced $2\pi$ microwave pulses ($\nu_{e1}$) to the electron spin. The position of the microwave pulses is schematically shown with vertical arrows. The nuclear spin nutation can then be released for further evolution to be locked later on in the opposite state (C) with a second burst of microwave pulses.

**5. Conclusions**

In this paper, we have reported our current progress in understanding spin relaxation for P donors in natural and $^{28}$Si-purified silicon. In natural Si the spin relaxation is strongly affected by spectral diffusion due to presence of 4.7% $^{29}$Si magnetic nuclei which decohere the electron spin in less than 0.5 ms. Very long relaxation times (extrapolating to $T_{2e}$ = 60 ms) have been found in isotopically pure $^{28}$Si. We have also demonstrated that the spin states of both the electron and nucleus of a $^{31}$P donor can be effectively controlled using resonant microwave and RF pulses. The *bang-bang* decoupling pulse protocol has been successfully implemented through the advantage of having coupled electron and nuclear spins in the donor. There are still many questions remaining to be answered. The implementation of two-qubit gates will require advanced processing, but the effects on spin coherence of this processing and of locating the spins in device structures is not yet known, for example. However, work on donor spins in silicon has established that this system can be considered a promising candidate for a future solid state quantum information processing technology.

## 6. Acknowledgements

We thank the Oxford-Princeton Link fund for support. Work at Princeton was supported by the ARO and ARDA under Contract No. DAAD19-02-1-0040. The research at Oxford is part of the QIP IRC www.qipirc.org (GR/S82176/01). GADB thanks EPSRC for a Professorial Research Fellowship (GR/S15808/01). AA and SCB are supported by the Royal Society. JJLM is supported by St. John's College, Oxford. Work at LBNL was supported by the Director, Office of Science, Office of Basic Energy Sciences, Materials Sciences and Engineering Division, of the U.S. Department of Energy under Contract No. DE-AC02-05CH11231.